\documentclass[prb,aps,twocolumn,showpacs,citeautoscript,longbibliography,
superscriptaddress]{revtex4-1}

\usepackage{amsmath}
\usepackage{bm}
\usepackage{amssymb}
\usepackage{graphicx}
\usepackage[colorlinks=true,allcolors=blue]{hyperref}

\begin{document}

\title{Universal power-law decay of electron-electron interactions due to
	nonlinear screening in a Josephson junction array}

\author{Daniel Otten}
\affiliation{JARA-Institute for Quantum Information, RWTH Aachen University,
D-52056 Aachen, Germany}

\author{Sebastian Rubbert}
\affiliation{Kavli Institute of Nanoscience, Delft University of Technology, P.O. Box 4056, 2600 GA Delft, The Netherlands}

\author{Jascha Ulrich}
\author{Fabian Hassler}

\affiliation{JARA-Institute for Quantum Information, RWTH Aachen University,
D-52056 Aachen, Germany}


\begin{abstract} 
  Josephson junctions are the most prominent nondissipative and at the same
  time nonlinear elements in superconducting circuits allowing Cooper pairs to
  tunnel coherently between two superconductors separated by a tunneling
  barrier.  Due to this, physical systems involving Josephson junctions show 
  highly complex behavior and interesting novel phenomena. Here, we
  consider an infinite one-dimensional chain of superconducting islands where
  neighboring islands are coupled by capacitances. We study the effect of
  Josephson junctions shunting each island to a common ground superconductor.
  We treat the system in the regime where the Josephson energy exceeds the
  capacitive coupling between the islands.  For the case of two offset charges
  on two distinct islands, we calculate the interaction energy of these charges
  mediated by quantum phase slips due to the Josephson nonlinearities.  We
  treat the phase slips in an instanton approximation and map the problem onto
  a classical partition function of interacting particles. Using the  Mayer
  cluster expansion, we find that the interaction potential of the offset
  charges decays with an universal inverse-square power law behavior.
\end{abstract}

\maketitle \section{Introduction}

In a circuit of capacitively coupled metallic islands, static screening
describes the redistribution of polarization charges on the capacitive plates
of the islands in response to a static offset charge on one of the islands.
The resulting voltages are determined by a (screened) Poisson equation. The
way in which the solution decays with the distance from the offset charge
depends on the effective circuit dimensionality.  In one-dimensional networks,
the polarization charge generically is constant up  to the screening length
and then follows a purely exponential decay. For metallic islands coupled by
capacitances $C$, the screening length $\sqrt{2 C/C_g}$ is determined by the
ratio of $C$ to the capacitances $C_g$ of the islands to
ground.\cite{fazio:01}

Josephson junctions in superconducting circuits add an interesting twist to
the screening of static offset charge configurations.  Conventional, linear
inductances coupling two metal grains are normally not of interest since they
invalidate the notion of islands with well-defined offset charges.  In
contrast, nonlinear Josephson inductances only allow tunneling of single
Cooper-pairs such that offset charge cannot simply flow off an island
contacted by a junction.  While Josephson junctions formally leave charge
quantization on the islands intact, charge quantization effects are
effectively weakened by large quantum fluctuations of the charge when the
charging energy $E_C$ of the islands is much smaller than the Josephson energy
$E_J$.  This gives rise to what we will in the following refer to as nonlinear
screening, an effect that has been exploited very successfully in the transmon
qubit~\cite{koch:07}.

In the regime of dominating Josephson energy, the dynamics of a single
junction is dominated by quantum phase slips corresponding to tunneling of the
superconducting phase difference by $2\pi$.  For one-dimensional chains of
Josephson junctions coupling the islands, the effects of phase slips have been
extensively studied theoretically both in infinite~\cite{bradley:84,
	korshunov:86, korshunov:89} and finite~\cite{matveev:02,
	Hekking13,ribeiro:14,vogt:15} networks in the past.  There has also
been considerable effort in studying these systems
experimentally~\cite{pop:10,manucharyan:12,ergul:13}.  Although many junctions
are present in these systems, the junctions are not strongly coupled such that
the dynamics are dominated by independent phase-slip events of the individual
junctions and interactions do not play a crucial role.

\begin{figure}
\includegraphics[scale=1.0]{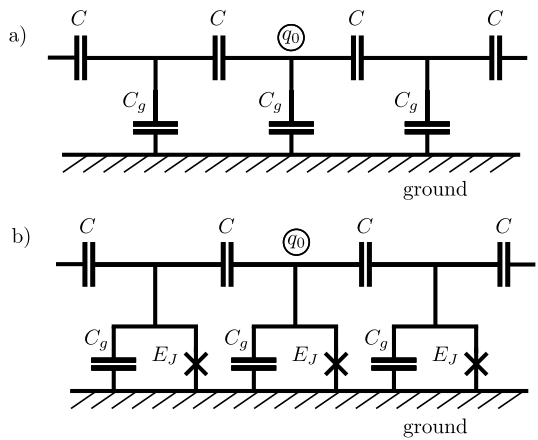}
\caption{\label{Fig:Setup} In a) we show a conventional one dimensional
infinite chain of capacitors with capacitance $C$. Each island is connected to
the  ground by a capacitance $C_g$. There is a fixed bias charge $q_0$ on a
single island, inducing charges on the neighboring capacitor plates. However,
due to the ground capacitances there is an exponential screening of this
charge along the chain. At each island a fraction of the induced charge is
stored on the ground capacitance so that the charge on the coupling
capacitances $C$ decays exponentially with the distance to the original bias
charge. The system in b) is similar to the system in a) with a Josephson
junction (Josephson energy $E_J$) providing an additional shunt to the ground.
Induced charges can now also be screened by tunneling through the Josephson
junction, giving rise to a novel nonlinear screening behavior. }
\end{figure}

While the nonlinear screening effected by a single Josephson junction as in
the transmon is well-studied, the screening properties of systems of many
junctions that are strongly coupled have not been investigated to the best of
our knowledge.  Motivated by the efficient screening of a single transmon, we
therefore study a one-dimensional system of transmons that are strongly
coupled by large capacitances $C$, see Fig.~\ref{Fig:Setup}.  This corresponds
to a system of superconducting islands that are coupled by capacitances $C$
and shunted to ground by a Josephson junction with Josephson energy $E_J$ and
associated capacitance $C_g$.  We are interested in a regime of nonlinear
screening dominated by the Josephson junctions which corresponds to a very
small capacitance $C_g$ to ground with associated large (linear) screening
length $\sqrt{2 C/C_g}$.  Phase slips dominate when the Josephson energy $E_J$
is much larger than the energy $E_{C} = e^2/2 C$ associated with a nonzero
voltage with respect to the ground on a single island.  We will show below
that the nonlinear screening due to the Josephson junctions leads to a
universal power-law decay of the electron-electron interaction with power two.
This implies a power-law decay of the polarization charge markedly different
from the conventional exponential decay obtained for static screening due to
capacitances.

The outline of the paper is as follows.  In Sec.~\ref{Sec:Setup}, we introduce
the problem and its corresponding imaginary-time partition function in the
path-integral formulation.  We discuss the dilute instanton-gas approximation
in Sec.~\ref{Sec:DiluteInstantonGas} and introduce the equal-time action that
is accumulated when slips of the island phases by $2\pi$ occur simultaneously
on different islands in Sec.~\ref{Sec:FirstOrderAction}.  In
Sec.~\ref{Sec:PartitionFunction}, we use the equal-time action to compute the
action of an arbitrary tunneling path of the island phases.  The corresponding
partition function maps onto a classical (interacting) partition function
which we compute using a Mayer expansion.  We use these results in
Sec.~\ref{Sec:GroundStateEnergy} to compute the ground state energy.  We
discuss the consequences for charge screening in
Sec.~\ref{Sec:ChargeScreening} and conclude with a short discussion of our
results.

\section{Setup and Model}
\label{Sec:Setup}

The system of interest is shown in Fig.~\ref{Fig:Setup}b). We analyze an
infinite one-dimensional chain of superconducting islands with the
superconducting phase $\varphi_j$ on the $j$-th island. The islands are
coupled by capacitances $C$ and connected to the ground by a Josephson
junction with Josephson energy $E_J$ and a capacitance $C_g$ in parallel. This
gives charges the possibility to tunnel on and off the island, changing the
screening behavior of the chain. We treat the problem within the quantum
statistical path integral approach to calculate the partition function $Z$ of
the system. As the phase-variables of the islands are compact and defined only
on the circle $[0,2\pi)$ it is useful to introduce the winding number $n_j \in
  \mathbb{Z}$ for the $j$-th island. With this, we can split the path integral
  for each phase $\varphi_j$ into sectors containing paths that wind $n_j$
  times around the circle. For the full partition function, we have to sum
  over all closed paths corresponding to all possible winding numbers.
  Additionally, we have to integrate over the starting positions $\phi_j$.
  Hence, the partition function is given by
\begin{align}
\label{Eq:PartitionSum}
Z=\prod_j\Biggl(\sum_{n_j}\int_0^{2\pi}\!d\phi_j\int_{\begin{subarray}{l} \varphi_j=\phi_j\end{subarray}}^{\varphi_f=\phi_j+2\pi n_j}\mathcal{D}[\varphi_j]\Biggr) e^{-S/\hbar},
\end{align}
where we have introduced the Euclidean action $S=\int_0^\beta\!
d\tau\;(L_C+L_q)$ with the inverse temperature $\beta=\hbar/k_B T$. The Lagrangian $L_C$ corresponding to the circuit without bias charges is given by
\begin{align}
L_C&=\sum_{j=-\infty}^{\infty} \Bigl\{ \frac{\hbar^2}{16 E_{C_g}} \dot{\varphi}_j^2+\frac{\hbar^2}{16 E_{C}}(\dot{\varphi}_{j+1}-\dot{\varphi}_j)^2\nonumber\\
&\quad-E_J[1-\cos(\varphi_j)] \Bigr \},
\end{align}
where $\dot\varphi_i=d\varphi_i/d\tau$. The first term in the sum describes
the capacitive coupling to the ground, the second term the coupling between
the islands, and the last term the Josephson junctions with Josephson energy
$E_J$ connecting the islands to the ground. The energy scales of the capactive terms are given by $E_{C_g}=e^2/2C_g$ and $E_{C}=e^2/2C$. To study the screening effect of the system in the presence of bias charges on selected islands, we need the additional Lagrangian 
\begin{align}
\label{Eq:BiasChargeLagrangian}
 L_q=\sum_{j=-\infty}^\infty \frac{i\hbar}{2e}q_j\dot{\varphi}_j
\end{align}
which implements the bias charges $q_j$ on the $j$-th island. This term is
special for two reasons: On one hand, it is a total time derivative and thus
does not enter the classical equations of motion. On the other hand, it is
imaginary so that it only adds a phase to the partition function underlining its nonclassicality.

From the free energy $F=-\hbar\log(Z)/\beta$, we can calculate the ground state energy $E$ of the system by applying the low temperature limit
\begin{align}
\label{Eq:LowTempLimit}
E=\lim_{\beta\rightarrow\infty} F.
\end{align}
The aim of this work is to calculate this ground state energy as a function of two bias charges and use it to gain information about the screening behavior of the chain.

\section{Partition function}\label{Sec:DiluteInstantonGas}

We are in particular interested in the regime where $E_{C}\ll E_J\ll E_{C_g}$ so
that the conventional capacitance to the ground $C_g$ is very small and the
Josephson junctions are mainly responsible for any charge screening on the
islands. From the fact that $E_J/E_{C}\gg 1$, we know that the ground state of
the system will be well-localized in the phase variables $\varphi_j$.
Therefore, the main contributions in (\ref{Eq:PartitionSum}) are due to paths
starting and ending in the minimum of the cosine potentials. As we are only
interested in exponential accuracy for the calculation of the ground state
energy with (\ref{Eq:LowTempLimit}), we can set $\phi_j=0$ and omit the
integral over $\phi_j$. We are left with the evaluation of 
 \begin{align}
\label{Eq:PartitionSum_0}
Z_0=\prod_j\Biggl(\sum_{n_j}\int_{\begin{subarray}{l} \varphi_j=0\end{subarray}}^{\varphi_f=2\pi n_j}\mathcal{D}[\varphi_j] \Biggr)e^{-S_C/\hbar+ i\pi\bm{n} \cdot \bm{q}/e}.
\end{align}
 Here, the action $S_C$ is $S_C = \int_0^\beta d \tau \, L_C$ and we have
 already carried out the time integral over the term
 \begin{align}
  \int_0^\beta\! d\tau \;L_q&=\frac{i\hbar}{2e}\sum_j q_j\int_0^\beta \!d\tau\;\dot{\varphi}_j
  =\frac{i\hbar\pi}{e}\; \bm{n} \cdot \bm{q}
 \end{align}
 due to the bias charges.
The vector $\bm{n}$ with components $n_j$ encodes the winding sector and $\bm{q}$ with components $q_j$ is the vector of bias charges.
As we are analyzing a regime where the phases are good variables, fluctuations
around the classical paths defined by the solutions of the Euler-Lagrange
equations (corresponding to $L_C$) are small. Hence, we apply an instanton
approximation where we replace the path integral by a sum over all classical
solutions, while quantum fluctuations around the classical paths play just a
sub-dominant role. In general, the main contribution to these fluctuations
arise from Gaussian integration of the action expanded to second order around
the classical paths. We assume that the fluctuations can be factorized  so
that they simply renormalize the bare parameters. This amounts to introducing
the weight prefactor $K(\bm{n},\varphi_\text{cl})$, accounting for the
fluctuations. By summing over all saddle point solutions of the Euler Lagrange
equations $\{\varphi_{\text{cl}}(\bm{n})\}$, the partition function in the instanton approximation reads    
 \begin{align}
 \label{Eq:PartitionSum_Instanton}
Z_{0}=\sum_{\bm{n}}\sum_{\{\varphi_{\text{cl}}(\bm{n})\}} K(\bm{n},\varphi_\text{cl}) e^{-S_{\text{cl}}[\varphi_{\text{cl}}]/\hbar+i\pi \bm{n} \cdot \bm{q}/e},
\end{align}
where $S_{\text{cl}}[\varphi_{\text{cl}}]$ is the action corresponding to the classical path $\varphi_{\text{cl}}(\bm{n})$. 
\begin{figure}
\includegraphics[scale=1.0]{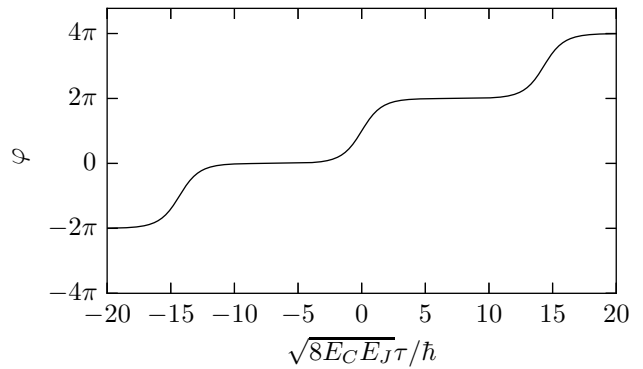}
\caption{\label{Fig:Path} A possible saddle-point solution of the of
equations of motion for a single island with phase variable $\varphi$. Every
step (phase slip) in the curve is described by an instanton particle localized in time. The action corresponding to such a path can be approximated by the action of a single phase slip times the number of total phase slips. This is possible because the constant parts of the steps do not contribute to the action. }
\end{figure}
An example of such a simple path for just a single island (only one $n_j$ is
different from 0) is shown in Fig.~\ref{Fig:Path}. 
The fact that we analyze the semi-classical regime where the phase is
well-localized allows  to use the dilute instanton gas approximation. The
approximation
holds as the phase-slip rate is so small that the phase slips (instantons) are well
separated from each other, i.e., there is at most a single phase slip present within
the duration $\tau_0=\hbar/\sqrt{E_J E_{C}}$ of a single phase-slip process. Thus, the classical paths
consist of almost instantaneous individual phase slips that are well-separated
in (imaginary) time. These phase slips are centered at their occurrence
times $\tau_j$ for the $j$-th phase slip. In between the phase slips, the phase stays
constant. In that way, we treat paths with more than a single phase slip at the
same island as independent phase slips, i.e., there is no temporal interaction
between the instantons. As a consequence the total action is simply the sum of
individual instanton contributions
\cite{Coleman85}. However, simultaneous phase slips at different islands cannot
be treated independently because they are subject to a spatial interaction due
to the coupling between the islands. Therefore, this case needs special
treatment that we deal with  in the next section. 

In principle we also have to calculate the prefactor $K(\bm{n})$ due to the
fluctuations around the classical action. These fluctuations are not
important when it comes to exponential accuracy. However, due to time
translation invariance, in the single instanton sector, the second-order
integration for the fluctuations additionally contains an integration of a zero
mode, corresponding to a simple shift of the full instanton solution in time.
We separate the prefactor $K(\bm{n})=\tilde{K}(\bm{n})\beta$ into a factor
$\tilde{K}(\bm{n})$ containing the real fluctuations on one side and the imaginary time interval
$\beta$ resulting from the zero mode integration on the other side. Thus, every
contribution is weighted by the fluctuations $\tilde{K}(\bm{n})$ and the length
of the time intervall $\beta$ in which we consider the evolution of the system.
For a single instanton on a single island, which is a noninteracting problem,
it is known that $\tilde{K}\simeq1/\tau_0$ (compare, e.g., to
Refs.~\onlinecite{koch:07,pop:10,matveev:02}). However, as the precision of
this prefactor is not as important as the precision for the exponentiated
instanton action we assume the single instanton value $1/\tau_0$ to be
sufficient even for the interacting problem with many simultaneous instantons
at different islands\cite{Sethna81}. This approximation is appropriate because
the terms become smaller with the number of instantons such that in the end
only prefactors  $\tilde{K}$ with a moderate amount of instantons are
relevant.
 
\section{Equal-Time Action}
\label{Sec:FirstOrderAction}
\begin{figure}
\includegraphics[scale=1.0]{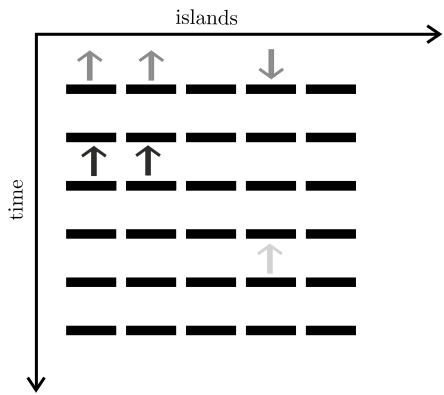}
\caption{\label{Fig:ConfigLattice}Example configuration of the system where
we have sliced the time dimension into a lattice to better visualize the finite
time an instanton process needs. The arrows pointing up correspond to
instantons, while the arrows pointing down correspond to anti-instantons.
(Anti-)instantons occurring at the same time are subject to a spatial
interaction (here marked by the same gray scale) and add a contribution to the
full action given by their equal-time action. For the full partition function,
we need to sum over all possible configurations. To that end, we consider each
phase slip event at an island as a particle with the island position,
occurrence time and instanton type as generalized coordinates. In this
picture, evaluating the full partition function corresponds to calculating the 
the classical grand-canonical partition function of the particles.} \end{figure}
Considering a single island with a single phase variable, the dilute instanton
gas approximation allows to treat the different phase slips (in imaginary
time) independently.  However, in our problem we have many interacting phase
degrees of freedom (in space) rendering the situation more involved.
Therefore, in this section, we determine the irreducible equal-time action
$S_{\text{ET}}$ including the simultaneous phase-slip processes explicitly.
For a proper definition of the equal-time action, we use the fact that within
a time interval of size $\tau_0$ there can be at most a single phase slip per
island.  Together with the diluteness of the instanton gas, it is convenient
to define the equal-time action as the action picked up by the total system in
a time window $\tau_0$ around a given time $\tau^*$. In this context it is
useful to imagine the (imaginary) time to be discrete with  a temporal lattice
constant $\tau_0$.  Figure~\ref{Fig:ConfigLattice} illustrates an example
configuration on such a lattice. For the calculation of $S_{\text{ET}}$ at the
time $\tau^*$, we only need the information which of the phases execute a
(anti-)phase slip. Later on, we will employ the equal-time action in the limit
of short instanton processes $\tau_0\rightarrow 0$, which applies in our
regime of interest, to construct the full action by adding the different
contributions independently in accordance with the dilute gas approximation.

In principle, for the explicit calculation of $S_{\text{ET}}$ at time $\tau^*$, we need to
extract the part of the classical paths matching the time window of size
$\tau_0$ around $\tau^*$ and insert this into the Lagrangian. However, as the
phase slips are almost instantaneous,  we are only interested whether a
particular island $j$ exhibits a phase slip ($n^*_j=1)$, an anti phase slip
($n^*_j=-1$) or no phase slip ($n^*_j=0$) at time $\tau^*$. Moreover, since the system does not pick up any action as long as the phases are constant,
we can extend the
time-integration from minus infinity to plus infinity as the phases are only nonconstant for the short time interval $\tau_0$ around
$\tau^*$. This yields $S_{\text{ET}}=\int_{-\infty}^{\infty}\!d\tau\; L_{C1}(\tau)$,
where $L_{C1}$ is the circuit Lagrangian with the classical solution for a
single phase slip per phase inserted. The boundary conditions are provided by
\begin{align}
\varphi_j(\tau)=2\pi \begin{cases}m_j^*,&\tau<\tau^*-\tau_0/2,\\
m_j^*+n_j^*,&\tau>\tau^*+\tau_0/2.
\end{cases}
\end{align}
Here, the discrete variable $m_j^*\in\mathbb{Z}$ contains the information
about the phase before the time interval of size $\tau_0$ around $\tau^*$.

The task is the calculation of the classical action for an interacting
nonlinear system that in general cannot be carried out exactly. Therefore, we
introduce an approximation to the nonlinear Josephson cosine potential by
replacing it by a periodic parabolic potential called the Villain
approximation\cite{Villain75}, i.e., $1-
\cos(\varphi_j)\approx \text{min}_{m_j}(\varphi_j-2\pi m_j)^2/2$ with
$m_j\in\mathbb{Z}$. Taking the minimum with respect to the discrete variable
$m_j$ corresponds to taking the phases modulo $2\pi$. In the case of no phase
slip with $n_j^*=0$ this means $m_j=m_j^*$ independent of the time. However, in
the case of $n_j^*=\pm1$ we have $m_j=m_j^*$ for $\tau<\tau^*$ and $m_j=m_j^*+n_j^*$ for $\tau >\tau^*$. This can be summarized for all cases as
\begin{align}
m_j(\tau)=m_j^*+n_j^*\Theta(\tau-\tau^*),
\end{align}
where $\Theta(\tau)$ is the Heaviside theta function. With the Euler-Lagrange
equations for the Villain potential, it is straightforward to show that the
Lagrangian $L_{C1}$ is mirror symmetric with respect to the time $\tau^*$.
Thus it is sufficient to calculate the action for times before $\tau^*$ and
double the result yielding $S_{\text{ET}}=2\int_{-\infty}^{\tau^*}\!d\tau
L_C(\tau)$. Additionally, the symmetry provides the boundary condition
$\varphi_j(\tau^*)=\pi (m_j^*+n_j^*)$.

Except from the bias charge term that does not change the
classical equations of motion, the system is translationally invariant and
thus can  be diagonalized by the Fourier transform
\begin{align}
\varphi_j&=\frac{1}{2\pi}\int_0^{2\pi}\! dk\; e^{i kj}\varphi_k,&
\varphi_k&=\sum_{j} e^{ -i kj}\varphi_j.
\end{align}  
Expressing the circuit Lagrangian $L_{C1}$ for $\tau\leq\tau^*$ in terms of $\varphi_k$ gives rise to
\begin{align}
L_{C1}=\frac{1}{2\pi}\int\!dk\;&\biggl\{\frac{\hbar^2}{16 E_{C_\Sigma} }\biggl[1-\frac{\cos(k)}{1+\varepsilon^2}\biggr]|\dot{\varphi}_k|^2\nonumber\\
&+E_J/2 |\varphi_k|^2 \biggr\},
\end{align}
where $E_{C_\Sigma}=e^2/2(C_g+2C)$ is the full charging energy and 
\begin{equation}
\varepsilon=
\sqrt{C_g/2C}
\end{equation}
the inverse screening length. At this point we make use of the fact that the Hamiltonian corresponding to $L_C$ is a conserved quantity. For the
instanton, it is equal to zero because the instantons correspond to
saddle-point solutions in the minima of the potentials. The conservation of
the Hamiltonian directly yields 
\begin{align}
\frac{\hbar^2}{16 E_{C_\Sigma} }\biggl[1-\frac{\cos(k)}{1+\varepsilon^2}\biggr]|\dot{\varphi}_k|^2=\frac{E_J}{2} |\varphi_k|^2.
\end{align}
With this equation, we can express the equal-time action as 
\begin{align}
S_{\text{ET}}(\bm{n}^*)&=2\int_{-\infty}^{\tau^*}\!d\tau\; L_{C1}\nonumber\\
&=\int\!\frac{dk}{\pi}\!\int_0^{\pi n_k^*}\!d|\varphi_k |\sqrt{\frac{E_J}{8E_{C_\Sigma}}\biggl[1-\frac{\cos(k)}{1+\varepsilon^2}\biggr]}|\varphi_k|\nonumber\\
&=\int \! dk\; U(k) |n_k^*|^2,
\end{align}
with $U(k)=\pi\sqrt{[1- \cos(k)/(1+\varepsilon^2)]E_J/32E_{C_\Sigma}}$ and $n_k^*$ the Fourier transform of $n_j^*$. In real space, we obtain the expression
\begin{align}
\label{Eq:RealSpaceAction}
S_{\text{ET}}(\bm{n}^*)&=\sum_{i,j}\; n_i^*  U(i-j) n_j^*,
\end{align}
where $U(j)$ is the Fourier transform of $U(k)$. For small $\varepsilon$ an accurate approximation for the real space potential can be given by 
\begin{align}
U(j)&=\alpha \frac{2\varepsilon j K_1(2\varepsilon j)}{\tfrac14-j^2}
\approx\alpha \frac{1}{\tfrac14-j^2} \quad(\text{for }\varepsilon\ll1).
\end{align}
Here, $K_1$ is the first modified Bessel function of the second kind with
$K_1(x)\approx 1/x$ for $x\ll1$. Hence the potential shows an inverse-square
decay until it reaches the screening length $\varepsilon^{-1}$ and turns into an exponential decay. The coupling strength is given by 
\begin{equation}
  \alpha=\pi\sqrt{E_J/8E_{C}}.
\end{equation}
Note that the equal-time action of interacting instantons can be fully described by the two-particle interaction $U(j)$ between all corresponding instantons. Additionally, we want to highlight that though the equal-time action describes the action picked up at a selected time it does not explicitly depend on the time but only on the underlying instanton configuration $\bm{n}$.

With the action at a given moment in time, we can proceed to calculate the
full action for a specific instanton configuration within the dilute gas
approximation.  In the next section, we are going to use the equal-time action
to calculate the partition function by summing over all instanton
configurations $\bm{n}$ and integrating over all times the instantons 
occur; this step is analogous to going over from a first to a second quantized
description of the problem.

\section{Ground State Energy}
\label{Sec:PartitionFunction}
We now turn to the estimation of the ground state Energy $E$. As a first step, we calculate the full partition function $Z_0$. With the instanton approximation and the equal-time action in real space, this is equivalent to a classical interacting statistical mechanics problem.
To make this correspondence clearer, we introduce a particle picture for the
phase slips. The general task is to evaluate
(\ref{Eq:PartitionSum_Instanton}) in the dilute gas approximation, which means
summing over all configurations of instantons on all islands
at all possible times. In the particle picture, the sum over all configurations
is realized by a sum over all numbers of instanton particles together with the
sum over the generalized coordinate $x_a$ of every particle $a=1,\dots,N$. The generalized coordinate $x_a$ of every particle includes its island coordinate $r_a \in
\mathbb{Z}$, the time $\tau_a$, and the instanton type $\sigma_a\in \{+,-\}$  (instanton or
anti-instanton) . Hence, such an instanton particle
corresponds to a single phase slip at a specific time and location.  We use the shorthand notation
\begin{align}
\int dx_a=\sum_{r_a,\sigma_a}\int_0^\beta\!\frac{d\tau_a}{\tau_0} \;
\end{align}
to express the summation over all configurations of the $a$-th particle. With
that, we can rewrite (\ref{Eq:PartitionSum_Instanton}) as
 \begin{align}
 \label{Eq:PartitionSum_Classical}
Z_{0}&=\sum_{N=0}^{\infty}\frac{z^N}{N!}\int \!dx_1\cdots dx_N\nonumber\\
&\qquad\quad
\times\exp\biggl[{-\sum_{\makebox[2.5em]{$\scriptstyle 1\leq a<b \leq N$}}V_{a,b}+i\sum_{1\leq a\leq N}\pi\sigma_a q(r_a)/e}\biggr],
\end{align}
which is a classical partition function in the grand canonical ensemble.
Note that the factor $N!$ prevents overcounting of the configurations. The
fugacity $z$ is defined by the self-interaction part of
Eq.~(\ref{Eq:RealSpaceAction}) (with $a=b$) of a single instanton with $z=\exp[{-U(0)}]$. In this context we can interpret $z/\tau_0$ as the instanton rate. As $z\ll1$ there will be much less than a single instanton per time $\tau_0$ on average, justifying the dilute gas approximation. The rest of the interacting part is absorbed in the interaction potential 
\begin{align}
V_{a,b}=\begin{cases}\infty, \hspace{2pt} &r_a=r_b,
  |\tau_a-\tau_b|\lesssim\tau_0/2 \\
2\sigma_a U(r_a-r_b)\sigma_b,\hspace{2pt}&r_a\neq r_b, |\tau_a-\tau_b|\lesssim
\tau_0/2\\
0,\hspace{5pt}&\text{else}.
\end{cases},
\end{align}
We implement the potential as a hardcore potential, so that only a single
phase slip can happen on a given island at a given time.
For phase slips occurring at different times, the interaction potential is zero because in the dilute gas approximation a spatial interaction between phase slips is only included for simultaneously events as explained in section \ref{Sec:FirstOrderAction}. The bias charge part is implemented by the single particle potential $\pi\sigma_a q(r_a)/e$, where $q(r)$  is the charge distribution over the islands. 

The free energy corresponding to such an interacting partition function can be evaluated perturbatively in $z$
by the Mayer cluster expansion \cite{Mayer41}. The idea is to rewrite
\begin{align}
\label{Eq:MayerTrick}
e^{-V_{a,b}}=1+f_{a,b},
\end{align}
so that we split the contribution in noninteracting part and interacting
part. The interacting part $f_{a,b}$ is, in the limit $\tau_0\rightarrow 0$,
proportional to a Dirac delta function \cite{Note1} $\tau_0\delta(\tau_a-\tau_b)$ with the width
$\tau_0$. This again reflects the fact that  in a dilute gas spatial
interaction affects only simultaneous instantons. For large distances
$|r_a-r_b|$, the interacting part $f_{a,b}$ is negligibly small, hence
suggesting an expansion in the number of interacting particles. We can proceed
similar with the bias charge potential. Here, we consider only two charges
separated by $M$ with the charge distribution
$q(r)=q_0\delta_{r,0}+q_M\delta_{r,M}$. We can write
\begin{align}
e^{i\pi\sigma_a q(r_a)/e}&=\exp({i\pi\sigma_a q_0 \delta_{r_a,0}/e})\exp({i\pi\sigma_a q_M \delta_{r_a,M}/e})\nonumber\\
&=(1+g_{0,a})(1+g_{M,a}),
\end{align}
where $g_{a,b}=\exp[i\pi\sigma_b q_a \delta_{r_b,a}/e]-1$ expresses the
interaction of phase slip $b$ with the charge on island $a$. Using these
relations, the partition function assumes the form
\begin{align}
\label{Eq:Mayer_PartitionSum}
Z_{0}=\sum_{N=0}^{\infty}\frac{z^N}{N!}&\int \!dx_1\cdots dx_N\nonumber\\
&\times\prod_{\begin{subarray}{c}a<b\\ l\end{subarray}} (1+f_{a,b})(1+g_{0,l})(1+g_{M,l}),
\end{align}
where the part in the product of the partition function contains terms with different numbers of $f$ functions like
\begin{align}
\prod_{a<b}(1+f_{a,b})&=[1+(f_{1,2}+f_{1,3}+\cdots)\nonumber\\
&\quad +(f_{1,2}f_{1,3}+f_{1,2}f_{1,4}+\cdots)+\cdots]
\end{align}
and similar for the $g$ functions.  A simple way to keep track the terms
appearing in the expansion is given by a diagrammatic approach: For every
particle coordinate $x_a$ in an $n$-particle term we draw a circle (node) with
the particle label $a$ inside. If an $f$-function $f_{a,b}$ is part of the
term, we connect the $a$-th and $b$-th circle by a straight line (link). A
$g_{0/M,l}$ is accounted for with a wiggled line starting from the $l$-th
circle and ending in a circle with the corresponding label $q_0$ or $q_M$. In
the end, we have to carry out a $dx_a$ integral for every node. Connected nodes represent interacting cluster of particles, which means that the integration of connected coordinates (clusters) is not necessarily independent, while nonconnected parts of the diagrams can be integrated independently. 

\begin{figure}
\includegraphics[scale=1.0]{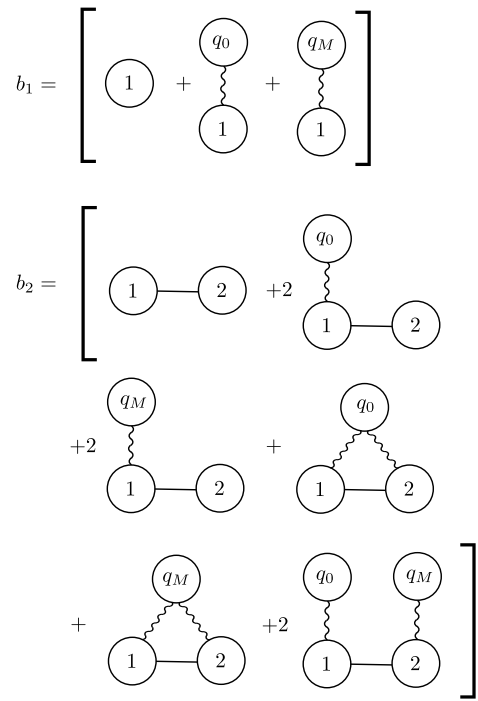}
\caption{\label{Fig:ClusterVars}The contributions to the cluster variables
$b_1$ and $b_2$ as given in Eq.~(\ref{Eq:ClusterVars}). The first-order
diagrams in $b_1$ contain no spatial interaction at all, while the last
diagram in $b_2$ mediates an interaction between the charges $q_0$ and $q_M$.
The factor of 2 in front of some of the diagrams is due to the fact that these
contributions can additionally be realized with the labels $1$ and $2$ interchanged. }
\end{figure}
It is important to realize that the value of such clusters after the
integration does not depend on their labels, but only on the cluster topology
and the number of coordinates included. Thus, we introduce the cluster variables $b_j$ given by
\begin{align}
\label{Eq:ClusterVars}
b_1&=\frac{1}{1!\,L}\int\! dx_1\;(1+g_{0,1}+g_{M,1}),\nonumber\\
b_2&=\frac{1}{2!\,L}\int \! dx_1dx_2 \;f_{1,2} \;[1+g_{0,1}+g_{M,1}+g_{0,2}+g_{M,2}\nonumber\\
&\qquad\qquad
+g_{0,1}g_{M,2}+g_{0,2}g_{M,1}+g_{0,1}g_{0,2}+g_{M,1}g_{M,2}],\nonumber\\
&\qquad\qquad\qquad\vdots
\end{align}
where $L$ is the number of islands in the system so that $b_n$ is a finite
quantity that includes all connected diagrams with $n$ particles and all their
possible interactions with the bias charges. Terms corresponding to a single
phase slip that are interacting with two charges at different islands do not
contribute and thus terms involving $g_{0,a}g_{M,a}$ vanish. In Fig.~\ref{Fig:ClusterVars} we show the diagrams corresponding to $b_1$ and $b_2$.

Every term in (\ref{Eq:Mayer_PartitionSum}) consists of different numbers of one,~two,~three and more-particle clusters. A term including $m_1$ single-particle clusters, $m_2$ two-particle clusters and so on contains $N$ particles with $N=\sum_j m_j j$. It contributes  
\begin{align}
T&=C(1!\,Lb_1)^{m_1}(2!\,Lb_2)^{m_2}(3!\,Lb_3)^{m_3}\cdots,\nonumber\\
C&=\frac{N!}{[(1!)^{m_1}(2!)^{m_2}\cdots][m_1!\,m_2!\cdots]}
\end{align}
where the Fa\`{a} di Bruno coefficient $C$ counts the number of ways of
partitioning the $N$ particles into the different particle clusters. The sum
over the instanton number $N$ in (\ref{Eq:Mayer_PartitionSum}) translates into
a sum over all cluster numbers $m_i$. Finally, we arrive at the expression for the partition function
\begin{align}
Z_0=\lim_{L\rightarrow\infty}\sum_{m_1,m_2,\dots}\biggl[\frac{(Lzb_1)^{m_1}}{m_1!}\frac{(Lz^2b_2)^{m_2}}{m_2!}\cdots\biggr]
\end{align}
and hence with (\ref{Eq:LowTempLimit}) the ground state energy is given by
\begin{align}
\label{Eq:FreeEnergyExpansion}
E=\lim_{\begin{subarray}{c}\beta\rightarrow\infty\\
  L\rightarrow\infty\end{subarray}}\Bigl(-\frac{L\hbar}{\beta}\sum_l b_l z^l
\Bigr).
\end{align}
This is an expansion in the fugacity $z$, where the order $l$ corresponds to the maximum number of particle clusters we take into account. Truncating the series at order 2 for example does not mean that we consider only terms with two instantons but that we only take interactions between pairs of instantons into account. In other words, we treat the system as a sum of many two-body problems. For small $z\ll1$, like in our case of $E_J/E_{C_\Sigma}\gg1$, such a truncation is justified. 

We analyze an infinite system and therefore the extensive energy $E$, scaling with the system size $L$, diverges. As we are interested in charge screening, we split the ground state energy 
\begin{align}
E=\mathcal{E}_0L+E_1(q_0)+E_1(q_M)+E_{2}(q_0,q_M),
\end{align}
where $\mathcal{E}_0$ is the energy density accounting for the bias charge
nonrelated energy per island that is stored in the chain. From the latter,
corresponding to all diagrams without a bias charge circle, we can in
principle get information about the pressure and other thermodynamic variables
in the system. However, here we are only interested in the influence due to
the bias charges. The parts including information about these break the
translational invariance of the chain and therefore do not scale with the
system size. This gives rise to the two energies $E_1(q)$ and $E_2(q_1,q_2)$,
where the first provides the change in the ground state energy due to a single charge and the latter corresponds to the interaction energy between two charges.

\section{Single Charge and Interaction Energy}\label{Sec:GroundStateEnergy}
We proceed by evaluating the ground state energy. First, we want to
calculate the ground state energy dependency on a single bias charge
corresponding to $E_1(q_0)$. In our case, it is enough to consider the first
order in (\ref{Eq:FreeEnergyExpansion}), because the second order is already
suppressed by an additional factor of the fugacity $z$. The task is to
calculate the part of $b_1$ corresponding to $E_1(q_0)$. The diagrammatic
expansion makes it easy to select the correct terms. There is only a single
diagram that we have to take into account: A single particle circle connected
to a single bias charge $q_0$ (Fig.~\ref{Fig:ClusterVars}, the second diagram in $b_1$), which is given by the simple expression 
\begin{align}
b_1(q_0)&=\frac{1}{1!\,L}\int\! dx_1\;g_{0,1},\nonumber\\
&=\frac{\beta}{L\tau_0}(e^{i\pi q_0/e}-1)+(e^{-i\pi q_0/e}-1)\nonumber\\
&=\frac{2\beta}{L\tau_0}[\cos(\pi q_0/e)-1].
\end{align}
As the $g$-functions are zero everywhere expect at the island where the
corresponding charge is located, they act as Kronecker-Delta and project the whole sum over the instanton coordinate on the location of the bias charge. From $b_1$, we find 
\begin{align}
\label{Eq:SingleEnergy}
E_1(q_0)&=\frac{2\hbar}{\tau_0} e^{-\pi\sqrt{2E_J/E_{C}}}[1-\cos(\pi q_0/e)].
\end{align}
This energy is similar to known results for problems with only single junctions
(see e.g.~\onlinecite{koch:07}). The exponent is slightly different from the
conventional scaling $\sqrt{8E_J/E_{C_g}}$ as every junction in the chain,
different from single junction systems, is coupled to other junctions. If we
redid the calculation in the limit $\varepsilon\rightarrow\infty$, which turns
off the coupling between the islands, and scale $E_J\mapsto (8/\pi^2)^2E_J$ to
compensate an error due to the Villain approximation \cite{Hekking13}, we
would recover the known result.

The next step is the calculation of the interaction energy $E_{2}(q_0,q_M)$.
Here, it is not enough to consider only first-order terms in
(\ref{Eq:FreeEnergyExpansion}), because single-instanton diagrams can only
contain single charges. Thus, the leading order of the interaction energy is
given by the two-instanton diagrams where every particle circle is connected
to a charge circle (the last contribution in $b_2$ in Fig.~\ref{Fig:ClusterVars}) corresponding to the expression
\begin{align}
b_2(q_0,q_M)&=\frac{1}{2!\,L}\int \! dx_1dx_2 \;2f_{1,2}g_{0,1}g_{M,2}.
\end{align}
By considering solely the terms depending on both charges in the result for $b_2(q_0,q_M)$,
we find the interaction energy
\begin{align}\label{eq:e2}
E_{2}(q_0,q_M)&=\frac{2\hbar}{\tau_0} e^{-2\pi\sqrt{2E_J/E_{C}}}\nonumber\\
&\quad\times \biggl\{\Bigl[e^{-2U(M)}-1\Bigr]\cos[\pi(q_0+q_M)/e]\nonumber\\
&\quad\quad+\Bigl[e^{2U(M)}-1\Bigr]\cos[\pi(q_0-q_M)/e]\biggr\}.
\end{align}
For large enough $M$, we have $U(M)\ll1$ and can therefore expand the
interaction energy in $U$ with the result 
\begin{align}
\label{Eq:InteractionEnergy}
E_{2}\approx \frac{8\hbar}{\tau_0}e^{-2\pi\sqrt{2E_J/E_{C}}}
U(M)\sin(\pi q_0/e)\sin(\pi q_M/e).
\end{align}
In this approximation, the charge interaction is directly proportional to the
instanton-interaction $U(M)$ and obeys a decay proportional to the
inverse-square distance between the charges (below the screening length).

\section{Charge Screening}\label{Sec:ChargeScreening}

In the final section, we return to the original question about the charge
screening effect of Josephson junctions. In a first step, we treat the case of
a single bias charge $q_0$ on island $0$. The presence of a charge on an
island induces an average charge on neighboring capacitor plates. To calculate
the latter we need to know the average voltages $V_M$ at the different
islands ($M\neq 0$). We can handle this task by using the results (\ref{Eq:SingleEnergy}) and (\ref{Eq:InteractionEnergy}) and employing linear response theory. The derivative of the ground state energy with respect to an external parameter
gives the average value of the derivative of the action with respect to the
same parameter. From (\ref{Eq:BiasChargeLagrangian}) and
(\ref{Eq:LowTempLimit}) and the time translation invariance in the system, we
obtain that the voltages $V_M$ are given by the derivative of the ground state
energy $E$ with respect to the bias charge. To this end, we shift the bias
charges on the $M$-th island such that the ground state energy
$E(q_0,\delta_M)$ depends on the small shift $\delta_M$.
We then find 
\begin{align}
\label{Eq:HellmanFeynman}
\frac{\partial E(q_0+\delta_0,\delta_M)}{\partial
\delta_M}\bigg|_{\delta_0,\delta_M=0}&=\frac{i\hbar}{2e}\langle
\dot\varphi_M\rangle = V_M;
\end{align}
here, in order to determine the voltages expectation values $V_M$ on island
$M$, we have used the Josephson relation $d\varphi_M/dt=2eV_M/\hbar$ together
with the relation $\tau = it$ between real and imaginary time.

\begin{figure}
\includegraphics[scale=1.0]{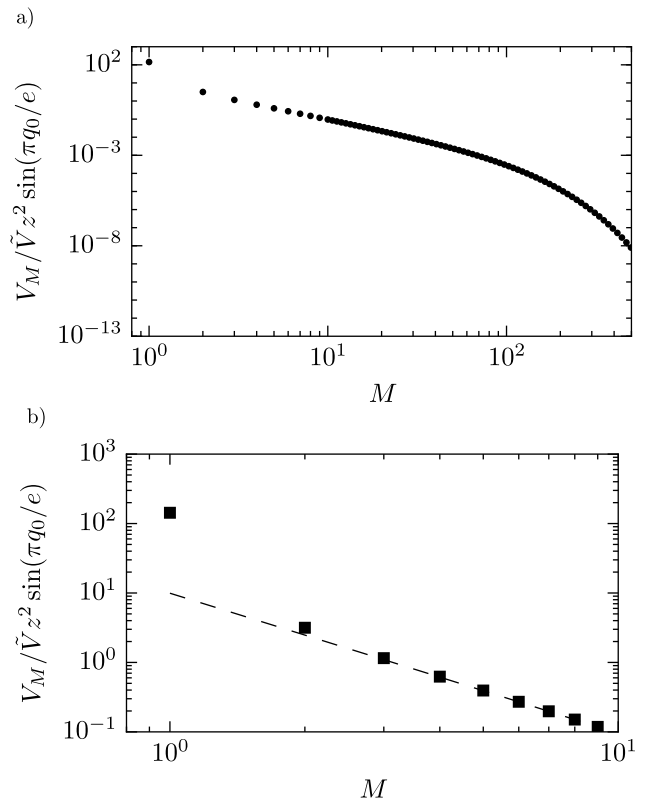}

\caption{\label{Fig:Charges}Double logarithmic plots of the voltage 
	$V_M$ induced on island $M$ by an offset charge at
	distance $M$ for  $E_J/E_{C}=5$ and $\varepsilon=0.01$. In panel a), we 
  show the full result
	corresponding to Eq.~\eqref{Eq:SeconOrderVoltage}. The screened voltage
  follows a power-law behavior until the screening length is reached at which point
  the
	exponential screening due to the ground capacitances $C_g$ takes over.
	In b) we compare the result obtained by using the power-law approximation
\eqref{Eq:ApproxVoltages} represented by the dashed line with the black
squares representing the full result. It can be
seen
that the decay of the induced voltages follows the inverse-square law  essentially starting from $M=2$.  }
\end{figure}
The leading contribution to $ E(q_0+\delta_0,\delta_M)$ is given by the
interaction energy
$E_2(q_0+\delta_0,\delta_M)$ of \eqref{eq:e2}. Taking the derivative, we
obtain the result ($\tilde{V}=2\pi\hbar/e\tau_0$)
\begin{align}
\label{Eq:SeconOrderVoltage}
V_M&=\tilde{V} e^{-2\pi\sqrt{2E_J/E_{C}}}
\Bigl[e^{-2U(M)}-e^{2U(M)}\Bigr]\sin(\pi q_0/e),
\end{align}
which can be approximated by the expression
\begin{align}
\label{Eq:ApproxVoltages}
V_M&\approx  -4\tilde{V} e^{-2\pi\sqrt{2E_J/E_{C}}} U(M)\sin(\pi q_0/e)
\nonumber\\
& \approx \frac{4 \tilde{V} \alpha
e^{-2\pi\sqrt{2E_J/E_{C}}} \sin(\pi q_0/e)}{M^2}  .
\end{align}
By expanding the exponential in (\ref{Eq:SeconOrderVoltage}) we see that all
even orders cancel so that there are no corrections until the third order in
$U(M)$. This makes the approximation (\ref{Eq:ApproxVoltages}) already
accurate for $M>\sqrt{2\alpha-1/4}$, where the exponent is smaller than 1.
Even in the regime of our interest $E_J\gg E_{C}$, $M$ does not have to be too
large as $\alpha$ scales with the square root of $E_J/E_{C}$.
Hence for intermediate distances (smaller than the screening length
$\varepsilon^{-1}$, larger than $\sqrt{2\alpha-1/4}$) the voltages obey a
universal
inverse-square decay given by $U(M)$. In Figure~\ref{Fig:Charges} we show a plot of the decay of the induced voltages. Note that an additional charge can be
treated by adding up the voltage contributions of every single charge.
Deviations from this simple rule are induced by a three particle
interaction-energy at least. Such contributions appear in three-instanton
diagrams or higher and thus they are strongly suppressed. For completeness, we
provide also the expression for the voltage on the $0$-th island 
\begin{align}
\label{Eq:FirstOrderVoltage}
V_0=\tilde{V} e^{-\pi\sqrt{2E_J/E_{C}}}\sin(\pi q_0/e),
\end{align}
 obtained
from $E\approx E_1(q_0+\delta_0)$.

With the voltages at hand, it is a simple task to determine the charges on all
capacitor plates. From the relation for capacitors $C(V_M-V_{M+1})=Q_M$, the
charge on the capacitor plate on the right of the $M$-th island (relative to
the bias charge) $Q_M$ is simply proportional to the voltage difference over
the capacitance. The largest voltage difference can be found between the bias
charge island and its two direct neighbors. Here we have to take the
difference of the first order contribution (\ref{Eq:FirstOrderVoltage}) and
the second-order contribution (\ref{Eq:SeconOrderVoltage}), where the latter
is suppressed by $z$. Thus, the two capacitor plates directly attached to the
bias charge island are charged the most. From the second island on, we only
need to take differences of (\ref{Eq:SeconOrderVoltage}). This results in the
power law  decay
\begin{equation}\label{eq:cubic}
  Q_M \approx \frac{8 C \tilde{V} \alpha
e^{-2\pi\sqrt{2E_J/E_{C}}} \sin(\pi q_0/e)}{M^3}.
\end{equation}
for $1\ll M \ll \epsilon^{-1}$.  The result is thus fundamentally different
from an exponential decay in usual linear screening.

\section{Conclusion}  
In this work, we have calculated the effect of Josephson junctions on the
charge screening in the ground state of an one-dimensional chain of
capacitively coupled superconducting islands in the semi-classical limit
$E_J/E_{C}\gg 1$. We have solved the problem of the interacting nonlinear
system by using an instanton approximation within the quantum statistical path
integral approach. To deal with the interactions in the chain, we have
introduced the equal-time action corresponding to the action picked up by the
whole system at a given moment in time. The latter includes spatial
correlations between simultaneous phase slips on different islands. With this
action and a dilute instanton gas approximation, which applies in the regime
of interest, we have mapped the task of solving the quantum system onto a
classical statistical mechanics problem. With a slightly modified Mayer
cluster expansion, supporting the interaction with bias charges at selected
islands of the chain, we have calculated a power series of the ground state
energy $E(q_0,q_M)$ in the number of interacting instantons as a function of
two bias charges $q_0$ and $q_M$. We  have calculated the average induced
voltages in the linear response regime and furthermore the induced charges on
the capacitor plates. Compared to the known exponential decay for chains
without the Josephson junctions, we have found that the induced voltages decay
with the inverse of the squared distance. This power-law decay is
fundamentally different from the conventional exponential screening. The
effect arises due to interacting quantum phase slips through the nonlinear Josephson potentials of the Josephson junctions.

\section{Acknowledgments} The authors acknowledge support from the Alexander
von Humboldt foundation and the Deutsche Forschungsgemeinschaft (DFG) under
grant HA 7084/2-1.

\end{document}